\documentclass[a4paper,fleqn]{cas-sc}

\usepackage{myStyle}
\usepackage{booktabs}
\usepackage{subcaption}
\newif\ifthesismode
\thesismodefalse 

\usepackage{setspace}
\usepackage{pgfornament}
\usepackage{eso-pic}
\AddToShipoutPictureBG*{%
	\AtPageUpperLeft{%
		\setlength\unitlength{1in}%
		\hspace*{\dimexpr0.5\paperwidth\relax} 
		\makebox(0,-0.5)[c]{\begin{tabular}{c c}
				Max van Haren \textit{et al}, Performance Analysis of Multirate Systems: A Direct Frequency-Domain Identification Approach, \\
				Mechanical Systems and Signal Processing 235:112843 (2025), DOI: \href{https://doi.org/10.1016/j.ymssp.2025.112843}{10.1016/j.ymssp.2025.112843}, uploaded June 2nd \\
		\end{tabular}}
}}

\begin{document}
	\let\WriteBookmarks\relax
	\def\floatpagepagefraction{1}
	\def\textpagefraction{.001}
	
\shorttitle{}
\shortauthors{Max van Haren et~al.}
	
	\title[mode = title]{Performance Analysis of Multirate Systems: A Direct Frequency-Domain Identification Approach}  
	
	
	
	\author[1]{Max van~Haren}[orcid=0000-0003-1243-1871]
	\cormark[1]
	\cortext[1]{Corresponding author.}
	\ead{m.j.v.haren@tue.nl}
	
	\author[2,1]{Lennart Blanken}[orcid=0000-0002-1892-2308]
	\author[1,3]{Tom Oomen}[orcid=0000-0001-7721-4566]
	\affiliation[1]{organization={Department of Mechanical Engineering, Control Systems Technology Section, Eindhoven University of Technology},
		addressline={Groene Loper 5},
		city={Eindhoven},
		postcode={5612 AE}, 
		country={The Netherlands}.}
\affiliation[2]{organization={Sioux Technologies},
	addressline={Esp 130},
	city={Eindhoven},
	postcode={5633 AA}, 
	country={The Netherlands}.}
	\affiliation[3]{organization={Delft Center for Systems and Control, Delft University of Technology},
		addressline={Mekelweg 2},
		city={Delft},
		postcode={2628 CN}, 
		country={The Netherlands}.}
	\begin{abstract}
		Frequency-domain performance analysis of intersample behavior in sampled-data and multirate systems is challenging due to the lack of a frequency-separation principle, and systematic identification techniques are lacking.
		The aim of this \manuscript is to develop an efficient technique for identifying the full intersample performance in the frequency-domain for closed-loop multirate systems, in particular the Performance Frequency Gain (PFG).
		Through local modeling techniques, aliased frequency components are effectively disentangled when identifying the PFG, which is directly facilitated by frequency-lifting the multirate system to a multivariable time-invariant representation.
		The developed method accurately and directly identifies the PFG in a single identification experiment.
		Finally, the developed method is experimentally validated on a prototype motion system, showing accurate identification of frequency-domain representations for the multirate system, including the PFG.
	\end{abstract}
\maketitle

\section{Introduction}
The performance of sampled-data systems is naturally defined in the continuous-time, i.e., the intersample performance. Unlike on-sample performance, which only considers performance at specific sampling instances, intersample performance additionally evaluates the system's behavior in between the sampling instances. Sampled-data systems include essentially all physical systems which are controlled by digital controllers \citep{Chen1995}, for example networked control systems \citep{Hespanha2007} and precision mechatronics \citep{Oomen2007}. The on-sample performance may vary significantly from the intersample performance, depending on the sampling time of the digital controller, the dynamics of the system, and the disturbances present. With this in mind, high-performance digital control designs should consider the intersample behavior of the system.


The intersample behavior of digital control systems can be considered through the use of sampled-data control techniques.
First, the controller can be designed in continuous-time and subsequently discretized \citep[Chapter~8]{Astrom2011}, where the digital control implementation is not considered during the design. 
Second, the system itself can be discretized in combination with a discrete-time control design, which ignores the intersample response of the system. 
Third, direct sampled-data control design \citep{Bamieh1991,Chen1995} overcomes these disadvantages by simultaneously considering the intersample response and the digital control implementation. 
Finally, multirate control design samples the system at an increased sampling rate relative to the controller's, which in combination with a discrete-time control design addresses the intersample response to a certain degree \textcolor{reviewblue}{\citep{Chen1995,Salt2005,Cimino2010}}.
However, both the first and second option do not consider the full intersample behavior, and the direct control approaches require an accurate continuous-time or fast-rate model, which is generally unknown or non-trivial to determine.


Effective frequency-domain modeling of sampled-data and multirate systems to capture the full intersample behavior is challenging, as their linear periodically time-varying nature \citep{Chen1995} prevents direct application of Linear Time-Invariant (LTI) methods due to signal aliasing. Therefore, alternative frequency-domain representations for sampled-data systems are developed in \citet{Araki1996,Yamamoto1996}, and include the Performance Frequency Gain (PFG) \citep{Lindgarde1997}. In addition, the PFG allows for an equivalent multirate definition \citep{Oomen2007}. The PFG captures the full intersample behavior of sampled-data and multirate systems, and can be effectively used for control design \citep{Oomen2007}. However, there is currently no efficient method to determine the PFG, as it either requires a continuous-time or fast-rate model of the system, which is generally unknown or difficult to identify, or an identification experiment for each input frequency, since aliasing is not accounted for.


Alternatively, the PFG can be identified indirectly through first identifying the underlying fast-rate system, followed by evaluating the PFG as described in \citet{Oomen2007}. The identification of the underlying fast-rate system can be done using either fast-rate outputs \Citep{VanHaren2022b} or downsampled outputs \Citep{VanHaren2025}. On the other hand, indirectly identifying the PFG using this two-step approach can lead to inaccurate results, and requires internal feedback signals that might be unavailable.


Although model-based sampled-data and multirate control design methods are broadly present, no effective and systematic frequency-domain identification techniques for these models are currently present. The aim of this \manuscript is to develop a fast, accurate, and inexpensive frequency-domain identification technique for closed-loop multirate systems that model the full intersample performance. The key idea in this \manuscript is to disentangle aliased frequency components for closed-loop multirate systems through local modeling techniques \citep{Pintelon2012,McKelvey2012}. Frequency lifting the multirate system to a multivariable time-invariant representation \citep{Zhang1997Maths,Bittanti2009} directly facilitates the application of local modeling techniques, which are originally developed for LTI systems. Furthermore, the multivariable time-invariant representations are used to directly compute the PFG, which can readily be used for intersample performance evaluation \textcolor{reviewblue}{in multirate control design, such as those in \citet{Salt2005,Cimino2010}}. The contributions include the following.
\ifthesismode
\begin{itemize}[align=left,	labelwidth=1.25cm, labelsep=0cm, leftmargin=1.25cm]
	\item[I.C.1.] \contributionAnchor{Contribution:PFGID:i}{I.C.1} The representation of the closed-loop PFG through the use of frequency-lifted time-invariant representations of multirate systems.
	\item[I.C.2.] \contributionAnchor{Contribution:PFGID:ii}{I.C.2} Effective single-experiment frequency-domain identification of these time-invariant representations through local modeling techniques, enabling direct evaluation of the PFG through \contributionRef{Contribution:PFGID:i}.
	\item[I.C.3.] \contributionAnchor{Contribution:PFGID:iii}{I.C.3} Validation of the developed framework on an experimental motion system.
\end{itemize}
\else
\begin{itemize}
	\item[C1)] The representation of the closed-loop PFG through the use of frequency-lifted time-invariant representations of multirate systems.
	\item[C2)] Effective single-experiment frequency-domain identification of these time-invariant representations through local modeling techniques, enabling direct evaluation of the PFG through C1.
	\item[C3)] Validation of the developed framework on an experimental setup.
\end{itemize}
\fi
This work extends \citet{VanHaren2022b} by being directly capable of computing the PFG for the multirate system, and in addition is suitable for systems with lightly-damped resonant dynamics.

\paragraph*{Notation:}
Signals sampled at a fast sampling rate are denoted by subscript $h$ and signals sampled at a slow sampling rate by subscript $l$. The $N$-points and $M$-points Discrete Fourier Transform (DFT) for finite-time fast-rate and slow-rate signals are respectively given by
\begin{equation}
	\label{PFGID:eq:DFT1}
		\begin{aligned}
			X_h(e^{j\omega_k \tsh}) &= \sum_{\dt=0}^{N-1} x_h(\dt) e^{-j\frac{2\pi \dt k}{N}}, &&			X_l(e^{j\omega_k \tsl}) &= \sum_{\ldt=0}^{M-1} x_l(\ldt) e^{-j\frac{2\pi \ldt k}{M}} 
		\end{aligned}
\end{equation}
with sampling times $\tsh$ and $\tsl$, discrete-time indices for fast-rate signals $\dt\in \mathbb{Z}_{[0,N-1]}$ and slow-rate signals $\ldt\in \mathbb{Z}_{[0,M-1]}$ with integers $\mathbb{Z}$ and $N,M$ the amount of data points of the fast-rate and slow-rate signals and frequency bin $k\in\mathbb{Z}_{[0,N-1]}$, which relates to the frequency grid
\begin{equation}
	\label{PFGID:eq:omegak}
	\omega_k=\frac{2\pi k}{N \tsh}=\frac{2\pi k}{M \tsl} \in [0,\wsh),
\end{equation}
with \textcolor{reviewblue}{fast-rate} sampling frequency $\wsh\textcolor{reviewblue}{=\frac{2\pi}{\tsh}}$ in rad/s. The sampling times of the slow-rate and fast-rate signals relate as $\tsl=\fac\tsh$, with downsampling factor $\fac\in\mathbb{Z}_{>0}$ \textcolor{reviewblue}{resulting in the slow-rate sampling frequency $\wsl=\frac{2\pi}{\tsl}=\wsh/\fac$}. Hence, the signal lengths relate as $N=\fac M$. 
\section{Problem Definition}
In this section, the problem is defined. The control setting is introduced, with its corresponding frequency-domain input-output analysis. Finally, the problem considered in this paper is defined.
\subsection{Control Setting}
\begin{figure}[tb]
	\centering
	\includegraphics{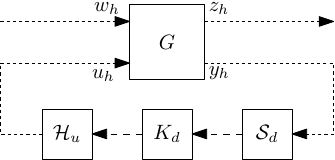}
	\caption{Multirate system, with fast-rate system $G$ and slow-rate controller $K_d$, which utilizes upsampler $\mathcal{H}_u$ and downsampler $\mathcal{S}_d$.}
	\label{PFGID:fig:MRCL_GeneralizedPlant}
\end{figure}

The control setting in \figRef{PFGID:fig:MRCL_GeneralizedPlant} is considered, where a fast-rate \textcolor{reviewblue}{LTI} system $G$ is under control with a slow-rate controller $K_d$. \textcolor{reviewblue}{The system $G$ is sampled at the fast sampling rate \wsh, whereas the controller $K_d$ is sampled at a reduced sampling rate $\wsl={\wsh}/{\fac}$.}
The exogenous signals $w_h$ contain any external signals, e.g., references or noise sources. The performance variable $z_h$ is for example the tracking error of the system. \textcolor{reviewblue}{Note that while the excitation signal $w_h$ and the performance variable $z_h$ are assumed to be known, they may contain noise.}
The \textcolor{reviewblue}{LTI} system $G$ is described by
\begin{equation}
	\begin{aligned}
		G = \left[\begin{array}{c|c}
			G_{11} & G_{12} \\ \hline
			\vphantom{\sum_{1}^{1}}G_{21} & G_{22}
		\end{array}\right].
	\end{aligned}
\end{equation}
\textcolor{reviewblue}{For ease of notation, the elements $G_{11}$, $G_{12}$, $G_{21}$, and $G_{22}$ are assumed single-input single-output throughout the paper. The notation can be straightforwardly extended to multivariable systems.}
The interpolator $\mathcal{H}_u$ consists of a zero-order hold filter and an upsampler $\mathcal{H}_u=\mathcal{I}_{\scriptscriptstyle ZOH}(q) \mathcal{S}_u$, with upsampler \citep{Vaidyanathan1993}
\begin{equation}
	\begin{aligned}
		\nu_h(\dt) = \mathcal{S}_u\nu_l(\ldt) = \begin{cases}
			\nu_l\left(\frac{\dt}{\fac}\right) & \text{for } \frac{\dt}{\fac}\in\mathbb{Z},\\[3pt]
			0& \text{for } \frac{\dt}{\fac}\notin\mathbb{Z}.
		\end{cases}
	\end{aligned}
\end{equation}
The zero-order hold filter is defined by
\begin{equation}
	\label{PFGID:eq:ZOH}
	\begin{aligned}
		\mathcal{I}_{\scriptscriptstyle ZOH}(q) = \sum_{f=0}^{\fac-1}q^{-f},
	\end{aligned}
\end{equation}
with fast-rate shift operator $q\nu_h\left(\dt\right) = \nu_h\left(\dt+1\right)$. The downsampler $\mathcal{S}_d$ is described by \citep{Vaidyanathan1993}
\begin{equation}
	\label{PFGID:eq:downsampler}
	\begin{aligned}
		\nu_l(\ldt) = \mathcal{S}_d\nu_h(\dt) = \nu_h(\fac \ldt).
	\end{aligned}
\end{equation}

\subsection{Frequency-Domain Analysis of Multirate System}
In this section, the frequency-domain behavior of multirate systems is described, and it is shown that the frequency-separation principle does not hold. First, after absorbing the feedback controller $\mathcal{H}_uK_d\mathcal{S}_d$ into the system $G$, the fast-rate input-output behavior of the closed-loop multirate system is described as
\begin{equation}
	\label{PFGID:eq:ClosedLoopIO}
	\begin{aligned}
		z_h =\left(G_{11} +  G_{12} \mathcal{H}_u \left(I-K_d\mathcal{S}_dG_{22} \mathcal{H}_u\right)^{-1}K_d\mathcal{S}_dG_{21}\right)w_h.
	\end{aligned}
\end{equation}
Taking the DFT \eqref{PFGID:eq:DFT1} on both sides of \eqref{PFGID:eq:ClosedLoopIO}, the output is described in the frequency-domain by \citep{Oomen2007}
\begin{equation}
	\label{PFGID:eq:ClosedLoopIO_DFT_PFG}
	\ifthesismode
	\begin{aligned}
		Z_h(e^{j\omega_k\tsh}) &= G_{11}\left(e^{j\omega_k\tsh}\right)W_h\left(e^{j\omega_k\tsh}\right) \\ &+G_{12}\left(e^{j\omega_k\tsh}\right)\mathcal{I}_{\scriptscriptstyle ZOH}\left(e^{j\omega_k\tsh}\right)Q_d\left(e^{j\omega_k\tsl}\right) \\
		&\cdot\frac{1}{\fac}\sum_{f=0}^{\fac-1}G_{21}\left(e^{j\left(\omega_{k}+\left(f/\fac\right)\wsh\right)\tsh}\right)W_h\left(e^{j\left(\omega_{k}+\left(f/\fac\right)\wsh\right)\tsh}\right),
	\end{aligned}
	\else
		\begin{aligned}
			Z_h(e^{j\omega_k\tsh}) &= G_{11}\left(e^{j\omega_k\tsh}\right)W_h\left(e^{j\omega_k\tsh}\right) \\ &+G_{12}\left(e^{j\omega_k\tsh}\right)\mathcal{I}_{\scriptscriptstyle ZOH}\left(e^{j\omega_k\tsh}\right)Q_d\left(e^{j\omega_k\tsl}\right) 			\frac{1}{\fac}\sum_{f=0}^{\fac-1}G_{21}\left(e^{j\left(\omega_{k}+\left(f/\fac\right)\wsh\right)\tsh}\right)W_h\left(e^{j\left(\omega_{k}+\left(f/\fac\right)\wsh\right)\tsh}\right),
		\end{aligned}
	\fi
\end{equation}
where
\begin{equation}
	\label{PFGID:eq:Qd}
	\begin{aligned}
		Q_d\!\left(e^{j\omega_k\tsl}\right) \!=\! \left(1\!-\!K_d\left(e^{j\omega_k\tsl}\right)\! G_{22,l}\left(e^{j\omega_k\tsl}\right)\right)^{-1}\!K_d\left(e^{j\omega_k\tsl}\right)\!.
	\end{aligned}
\end{equation}
The slow-rate system $G_{22,l} = \mathcal{S}_dG_{22}\mathcal{H}_u$ in \eqref{PFGID:eq:Qd} is described in the frequency domain as \citep{Vaidyanathan1993}
\begin{equation}
	\label{PFGID:eq:DownsampledFRF}
	\begin{aligned}
		G_{22,l}\left(e^{j\omega_k\tsl}\right) = 	\frac{1}{\fac} \sum_{f=0}^{\fac-1}\left( G_{22}\left(e^{j\left(\omega_{k}+\left(f/\fac\right) \wsh\right)\tsh}\right)\cdot\mathcal{I}_{\scriptscriptstyle ZOH}\left(e^{j\left(\omega_{k}+\left(f/\fac\right) \wsh\right)\tsh}\right)\right).
	\end{aligned}
\end{equation}

A key observation is that the intersample behavior $Z_h\left(e^{j\omega_k\tsh}\right)$ is influenced by \fac frequencies of the input $W_h\left(e^{j\left(\omega_k+(f/\fac)\wsh\right)\tsh}\right)$ due to aliasing. Conversely, each frequency of the input $W_h\left(e^{j\omega_k\tsh}\right)$ influences \fac frequencies of the output \textcolor{reviewblue}{\citep[Theorem~3]{Salt2014}}. This dependency makes it unclear how to analyze the frequency-domain intersample behavior for multirate systems. As a result, there is a need for a systematic approach for analyzing this behavior, particularly in a way that is useful for control design.

\subsection{Problem Definition}
The problem considered in this \manuscript is as follows. Given a fast-rate excitation signal $w_h$ and performance variable $z_h$ of the multirate system shown in \figRef{PFGID:fig:MRCL_GeneralizedPlant}, directly identify the relevant frequency-domain representations that include the full intersample performance, i.e., the PFG.

\section{Method}
\ifthesismode
	In this section, a frequency-domain representation of the multirate system is introduced through time-invariant representations, constituting \contributionRef{Contribution:PFGID:i}. Furthermore, the time-invariant representation is identified in a single identification experiment through local modeling, leading to \contributionRef{Contribution:PFGID:ii}. 
\else
	In this section, a frequency-domain representation of the multirate system is introduced through time-invariant representations, constituting contribution C1. Furthermore, the time-invariant representation is identified in a single identification experiment through local modeling, leading to contribution C2. 
\fi
The developed approach is then summarized in a procedure.
\subsection{Intersample Performance Analysis through the PFG}
An effective frequency-domain approach for analyzing the intersample performance for the multirate system in \figRef{PFGID:fig:MRCL_GeneralizedPlant} is the PFG, which analyzes the total power output for a single-frequency input. The PFG is given by
\begin{equation}
	\label{PFGID:eq:PFGDef}
	\begin{aligned}
		\mathcal{P}\left(e^{j \omega_d \tsh}\right)\underset{w_h \in \mathcal{W}}{=} \frac{\left\|z_h\right\|_{\mathcal{P}}}{\left\|w_h\right\|_{\mathcal{P}}},
	\end{aligned}
\end{equation}
where the signal space $\mathcal{W}$ consists of single complex sinusoidal disturbances \textcolor{reviewblue}{having frequency $\omega_d$ and amplitude $c$}, i.e.,
\begin{equation}
	\label{PFGID:eq:signalSpace}
	\begin{aligned}
		\mathcal{W}=\Big\{w_h(\dt) | w_h(\dt)=c e^{j \omega_{\textcolor{reviewblue}{d}} \dt \tsh},0<\|c\|_2<\infty\Big\},
	\end{aligned}
\end{equation}
The power $\|x_h\|_\mathcal{P}$ in \eqref{PFGID:eq:PFGDef} is given by
\begin{equation}
	\label{PFGID:eq:Power}
	\begin{aligned}
		\|x_h\|_{\mathcal{P}}=\sqrt{\lim _{N \rightarrow \infty} \frac{1}{N} \sum_{\dt=0}^{N-1}\left\|x_h(\dt)\right\|^2_2}.
	\end{aligned}
\end{equation}
Alternatively, the PFG is calculated in the frequency-domain as presented in \lemmaRef{lemma:fPFG}.
\begin{lemma}
	\label{lemma:fPFG}
	The PFG \eqref{PFGID:eq:PFGDef} is equivalent to
	\begin{equation}
		\label{PFGID:eq:PFGFreqDomainDef}
		\begin{aligned}
			\mathcal{P}\left(e^{j \omega_d \tsh}\right)\underset{w_h \in \mathcal{W}}{=} \frac{\left\|Z_h\right\|_{\mathcal{P}}}{\left\|W_h\right\|_{\mathcal{P}}},
		\end{aligned}
	\end{equation}
	with 
	\begin{equation}
		\label{PFGID:eq:FrequencyDomainPower}
		\begin{aligned}
			\|X_h\|_{\mathcal{P}}=\sqrt{\lim _{N \rightarrow \infty} \frac{1}{N} \sum_{k=0}^{N-1}\left\|X_h\left(e^{j\omega_k \tsh}\right)\right\|_2^2}.
		\end{aligned}
	\end{equation}
\end{lemma}
\begin{proof}
	From Parseval's theorem it is known that
	\begin{equation}
		\begin{aligned}
			\sum_{\dt=0}^{N-1} \left\|x\left(\dt\right)\right\|_2^2 = \frac{1}{N} \sum_{k=0}^{N-1}\left\|X\left(e^{j\omega_k \tsh}\right)\right\|_2^2,
		\end{aligned}
	\end{equation}
which, when substituted into \eqref{PFGID:eq:Power}, and subsequently in \eqref{PFGID:eq:PFGDef} for $w_h$ and $z_h$ directly leads to \eqref{PFGID:eq:PFGFreqDomainDef}.
\end{proof}

Note that an input signal $w_h\in\mathcal{W}$ \eqref{PFGID:eq:signalSpace} consists of a single sinusoid, and hence results in a DFT \eqref{PFGID:eq:DFT1} magnitude of
\begin{equation}
	\label{PFGID:eq:SignalSpaceDFT}
	\begin{aligned}
		\left\|W_h\left(e^{j\omega_k \tsh}\right)\right\|_2\underset{w_h \in \mathcal{W}}{=}\begin{cases}
			cN & \text{ for } \omega_k=\omega_{\textcolor{reviewblue}{d}}, \\
			0 & \text{ otherwise}.
		\end{cases}
	\end{aligned}
\end{equation}
Therefore, the power of the input signal is determined by substituting \eqref{PFGID:eq:SignalSpaceDFT} in \eqref{PFGID:eq:FrequencyDomainPower}, resulting in \begin{equation}
	\label{PFGID:eq:inputPower}
	\begin{aligned}
		\left\|W_h\right\|_\mathcal{P}\underset{w_h \in \mathcal{W}}{=}c\sqrt{N}.
	\end{aligned}
\end{equation}

The PFG \eqref{PFGID:eq:PFGDef} represents the full intersample behavior of the multirate system in \figRef{PFGID:fig:MRCL_GeneralizedPlant}, since it takes into account all output frequencies, including aliased ones, for a single input frequency. Due to the aliasing of signals in \eqref{PFGID:eq:ClosedLoopIO_DFT_PFG}, identifying the PFG is time-consuming since the excitation signal $w_h$ is limited to a single frequency.

\subsection{Direct PFG Identification through Frequency-Lifting}
\begin{figure}[tb]
	\centering
	\includegraphics{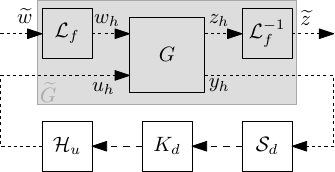}
	\caption{Frequency-lifted multirate system, where frequency-lifting operators $\mathcal{L}_f$ and $\mathcal{L}_f^{-1}$ transform the multirate system to a time-invariant multivariable system.}
	\label{PFGID:fig:MRCL_GeneralizedPlantfLifted}
\end{figure}

The multirate PFG is directly identified through frequency-lifting the multirate system to a multivariable time-invariant representation. The frequency-lifted signal $\widetilde{X}\left(e^{j\omega_k\tsh}\right) = \mathcal{L}_f {X}\left(e^{j\omega_k\tsh}\right)$ is defined as
\begin{equation}
	\label{PFGID:eq:fLiftedSignal}
	\begin{aligned}
		\widetilde{X}\left(e^{j\omega_k\tsh}\right) = \begin{bmatrix}
			{X}\left(e^{j\omega_k\tsh}\right) \\
			{X}\left(e^{j\omega_k\tsh} \phi \right) \\
			\vdots\\[0.5mm]
			{X}\left(e^{j\omega_k\tsh}\phi^{\fac-1}\right)
		\end{bmatrix} \in\mathbb{C}^{\fac},
	\end{aligned}
\end{equation}
where $\phi=e^{j2\pi/\fac}$ corresponds to a frequency shift of $\wsh/\fac$ rad/s. Note that therefore the $i^{\textrm{th}}$ entry of $\widetilde{X}$ is essentially the original signal shifted in frequency by $i\wsh/\fac$. By frequency-lifting the exogenous inputs $\widetilde{w}=\mathcal{L}_fw_h$ and performance outputs $\widetilde{z}=\mathcal{L}_fz_h$, the system becomes
\begin{equation}
	\begin{aligned}
		\widetilde{G} = \left[\begin{array}{c|c}
			\mathcal{L}_f G_{11} \mathcal{L}_f^{-1} & \mathcal{L}_f G_{12} \\ \hline
			  \vphantom{\Big(}G_{21}\mathcal{L}_f^{-1} & G_{22}
		\end{array}\right].
	\end{aligned}
\end{equation}
The frequency-lifted system is shown in \figRef{PFGID:fig:MRCL_GeneralizedPlantfLifted}.
By absorbing the feedback controller $\mathcal{H}_uK_d\mathcal{S}_d$ into the interconnection, the closed-loop transfer describing the frequency-lifted input-output behavior is
\begin{equation}
	\label{PFGID:eq:NTilde}
	\begin{aligned}
		\widetilde{M} = &\mathcal{L}_f G_{11} \mathcal{L}_f^{-1} + \mathcal{L}_f G_{12} \mathcal{H}_uK_d\mathcal{S}_d \left(I-G_{22} \mathcal{H}_uK_d\mathcal{S}_d\right)^{-1}G_{21}\mathcal{L}_f^{-1},
	\end{aligned}
\end{equation}
which is LTI and multivariable $\widetilde{M}\left(e^{j\omega_k\tsh}\right)\in\mathbb{C}^{\fac\times\fac}$ \citep{Bittanti2009}. Therefore, the DFT of the input-output behavior is described by
\begin{equation}
	\label{PFGID:eq:fLiftedIO}
	\begin{aligned}
		\widetilde{Z}\left(e^{j\omega_k\tsh}\right) = \widetilde{M}\left(e^{j\omega_k\tsh}\right)\widetilde{W}\left(e^{j\omega_k\tsh}\right),
	\end{aligned}
\end{equation}
where $\widetilde{M}\left(e^{j\omega_k\tsh}\right)\in\mathbb{C}^{\fac\times\fac}$.

The frequency-lifted closed-loop $\widetilde{M}$ is directly related to the PFG of the multirate system, which leads to the main result in this section in \theoremRef{theorem:fLiftedPFG}.
\begin{theorem}
	\label{theorem:fLiftedPFG}
	The PFG \eqref{PFGID:eq:PFGDef} of closed-loop multirate system in \figRef{PFGID:fig:MRCL_GeneralizedPlant} for frequencies \textcolor{reviewblue}{$\omega_d\in\left[0,\wsh\right)$} is equivalent to
	\begin{equation}
		\label{PFGID:eq:fLiftedPFGCalc}
		\begin{aligned}
			\mathcal{P}\left(e^{j\omega_{\textcolor{reviewblue}{d}}\tsh}\right) = \sqrt{\sum_{f=0}^{\fac-1}\left\|\widetilde{M}_{[f+1,1]}\left(e^{j\omega_{\textcolor{reviewblue}{d}}\tsh}\right)\right\|_2^2},
		\end{aligned}
	\end{equation}
where $X_{[i,j]}\left(e^{j\omega_k\tsh}\right)$ denotes the $(i,j)^{\textrm{th}}$ element of matrix $X$.
\end{theorem}
\begin{proof}
First, $\left\|Z_h\right\|_\mathcal{P}$ in the PFG \eqref{PFGID:eq:PFGFreqDomainDef} is computed by splitting the sum in \eqref{PFGID:eq:FrequencyDomainPower} into \fac frequency bands, and represented through the use of the frequency-lifted outputs as
\begin{equation}
	\label{PFGID:eq:SplitSum}
	\begin{aligned}
		\ifthesismode
			\sum_{k=0}^{N-1}\left\|Z_h\left(e^{j\omega_k\tsh}\right)\right\|_2^2 = \!\!\sum_{k=0}^{M-1}\sum_{f=0}^{\fac-1} \left\|Z_h\left(e^{j\omega_k\tsh}\phi^f\right)\right\|_2^2 = \!\!\sum_{k=0}^{M-1} \sum_{f=0}^{\fac-1} \left\|\widetilde{Z}_{[f+1]}\left(e^{j\omega_k\tsh}\right)\right\|_2^2\!\!.
		\else
			\sum_{k=0}^{N-1}\left\|Z_h\left(e^{j\omega_k\tsh}\right)\right\|_2^2 = \sum_{k=0}^{M-1}\sum_{f=0}^{\fac-1} \left\|Z_h\left(e^{j\omega_k\tsh}\phi^f\right)\right\|_2^2 = \sum_{k=0}^{M-1} \sum_{f=0}^{\fac-1} \left\|\widetilde{Z}_{[f+1]}\left(e^{j\omega_k\tsh}\right)\right\|_2^2.
		\fi
	\end{aligned}
\end{equation}
Second, under excitation signal $w_h\in\mathcal{W}$ with magnitude \eqref{PFGID:eq:SignalSpaceDFT} and utilizing \eqref{PFGID:eq:fLiftedIO} the magnitude of these frequency-lifted outputs are 
\begin{equation}
	\label{PFGID:eq:ZtildeEntries}
		\begin{aligned}
\left\|\widetilde{Z}_{[f+1]}\left(e^{j\omega_k\tsh}\right)\right\|_2 \underset{w_h\in\mathcal{W}}{=}  \begin{cases}
					cN\left\|\widetilde{M}_{[f+1,1]}\left(e^{j\omega_k\tsh}\right)\right\|_2, & \text{for } \omega_k = \omega_{\textcolor{reviewblue}{d}}. \\
					0, & \text{otherwise}.
			\end{cases}
		\end{aligned}
\end{equation}
By substitution of \eqref{PFGID:eq:SplitSum} and \eqref{PFGID:eq:ZtildeEntries} in \eqref{PFGID:eq:FrequencyDomainPower}, $\left\|Z_h\right\|_\mathcal{P}$ is formulated as
\begin{equation}
	\label{PFGID:eq:Zh_LiftedExpression}
	\begin{aligned}
		\left\|Z_h\right\|_\mathcal{P} \underset{w\in\mathcal{W}}{=} c\sqrt{N}\sqrt{\sum_{f=0}^{\fac-1}\left\|\widetilde{M}_{[f+1,1]}\left(e^{j\omega_{\textcolor{reviewblue}{d}}\tsh}\right)\right\|_2^2}.
	\end{aligned}
\end{equation}
Finally, substitution of \eqref{PFGID:eq:Zh_LiftedExpression} and \eqref{PFGID:eq:inputPower} in \eqref{PFGID:eq:PFGFreqDomainDef} leads to the main result \eqref{PFGID:eq:fLiftedPFGCalc}.
\end{proof}

\begin{remark}
	Note that while \theoremRef{theorem:fLiftedPFG} utilizes the first column of $\widetilde{M}\left(e^{j\omega_k\tsh}\right)$, any $f^{\textrm{th}}$ column can be used as well by frequency shifting its result with $\phi^{-f}$.
\end{remark}

By direct identification of the PFG as shown in \theoremRef{theorem:fLiftedPFG}, the frequency-lifted representation directly allows for intersample performance evaluation.

\subsection{Multivariable Identification through Local Modeling}
\label{sec:LocalModeling}
In this section, the frequency-lifted system $\widetilde{M}$ is effectively identified in a single identification experiment by disentangling aliased frequency components through multivariable local modeling techniques.

In a local frequency window $r\in\mathbb{Z}_{[-\wsize,\wsize]}$, the frequency-lifted output $\widetilde{Z}$ in \eqref{PFGID:eq:fLiftedIO} is approximated as
\begin{equation}
	\label{PFGID:eq:EstOutput}
	\begin{aligned}
		\widehat{\widetilde{Z}}\left( e^{j\omega_{k+r} \tsh}\right)  \!=\! \widehat{\widetilde{M}}\!\left( e^{j\omega_{k+r} \tsh}\right)\widetilde{W}\left( e^{j\omega_{k+r} \tsh}\right) \!+\! \widehat{\widetilde{T}}\left( e^{j\omega_{k+r} \tsh}\right)\\
	\end{aligned}
\end{equation}
where $\widehat{\widetilde{M}}\left( e^{j\omega_{k+r} \tsh}\right)\in\mathbb{C}^{\fac\times\fac}$ approximates the closed-loop $\widetilde{M}$ in \eqref{PFGID:eq:NTilde}, and transient term $\widehat{\widetilde{T}}\left( e^{j\omega_{k+r} \tsh}\right) \in\mathbb{C}^{\fac}$, which is introduced due to finite-length signals, and includes leakage effects. The multivariable system $\widehat{\widetilde{M}}$ and transient $\widehat{\widetilde{T}}$ are modeled using the local models
\begin{equation}
	\label{PFGID:eq:localModel}
	\ifthesismode
	\begin{aligned}
		\widehat{\widetilde{M}}\left( e^{j\omega_{k+r} \tsh}\right)  &= D^{-1}\left( e^{j\omega_{k+r} \tsh}\right)N\left( e^{j\omega_{k+r} \tsh}\right), \\
		\widehat{\widetilde{T}}\left( e^{j\omega_{k+r} \tsh}\right)  &= D^{-1}\left( e^{j\omega_{k+r} \tsh}\right)L\left( e^{j\omega_{k+r} \tsh}\right),
	\end{aligned}
	\else
	\begin{aligned}
		\widehat{\widetilde{M}}\left( e^{j\omega_{k+r} \tsh}\right)  = D^{-1}\left( e^{j\omega_{k+r} \tsh}\right)N\left( e^{j\omega_{k+r} \tsh}\right), &&
		\widehat{\widetilde{T}}\left( e^{j\omega_{k+r} \tsh}\right)  = D^{-1}\left( e^{j\omega_{k+r} \tsh}\right)L\left( e^{j\omega_{k+r} \tsh}\right),
	\end{aligned}
\fi
\end{equation}
\textcolor{reviewblue}{where local lifted system numerator $N\left(e^{j\omega_{k+r} \tsh}\right)\in\mathbb{C}^{\fac\times\fac}$, denominator $D\left(e^{j\omega_{k+r} \tsh}\right)\in\mathbb{C}^{\fac\times\fac}$, and transient numerator $L\left(e^{j\omega_{k+r} \tsh}\right)\in\mathbb{C}^{\fac}$ are given by}
\begin{equation}
	\label{PFGID:eq:LocalDenNum}
	\begin{aligned}
		N\left( e^{j\omega_{k+r} \tsh}\right) &= \widehat{\widetilde{M}}\left( e^{j\omega_{k} \tsh}\right) +\sum_{s=1}^{R_n}N_s(k)r^s, \\
		L\left( e^{j\omega_{k+r} \tsh}\right) &= \widehat{\widetilde{T}}\left( e^{j\omega_{k} \tsh}\right) +\sum_{s=1}^{R_l}L_s(k)r^s, \\
		D\left( e^{j\omega_{k+r} \tsh}\right) &= I +\sum_{s=1}^{R_d}D_s(k)r^s, \\
	\end{aligned}
\end{equation}
\textcolor{reviewblue}{with complex coefficients $N_s(k)\in\mathbb{C}^{\fac\times\fac}$, $L_s(k) \in\mathbb{C}^{\fac}$, and $D_s(k)\in\mathbb{C}^{\fac\times\fac}$.}
The decision parameters
\begin{equation}
	\begin{aligned}
			{\Theta}\left(k\right): \begin{cases}
				\widehat{\widetilde{M}}\left( e^{j\omega_{k} \tsh}\right) &\in \mathbb{C}^{\fac\times \fac}, \\
				\widehat{\widetilde{T}}\left( e^{j\omega_{k} \tsh}\right) &\in \mathbb{C}^{\fac}, \\
				N_s(k)&\in\mathbb{C}^{n_u\fac+n_y\times n_u\fac}, \\
				L_s(k)&\in\mathbb{C}^{n_u\fac+n_y}, \\
				D_s(k)&\in\mathbb{C}^{n_u\fac+n_y\times n_u\fac+n_y},
			\end{cases}
	\end{aligned}
\end{equation}
are determined by minimizing the weighted difference between approximated outputs \eqref{PFGID:eq:EstOutput} and measured outputs $Z\left( e^{j\omega_{k+r} \tsh}\right)$, resulting in the linear least squares problem
\begin{equation}
	\label{PFGID:eq:CostFunction}
	\ifthesismode
		\begin{aligned}
			\widehat{\Theta}(k) = \arg\min_{\Theta(k)} \sum_{r=- \wsize}^\wsize \Bigg\|
			&D\left( e^{j\omega_{k+r} \tsh}\right)\left(\widetilde{Z}\left( e^{j\omega_{k+r} \tsh}\right)-\widehat{\widetilde{Z}}\left( e^{j\omega_{k+r} \tsh}\right)\right)\Bigg\|_2^2 \\
			= \arg\min_{\Theta(k)} \sum_{r=- \wsize}^{\wsize} \Bigg\|\vphantom{\sum_{r=- \wsize}^{\wsize}}
			&D\left( e^{j\omega_{k+r} \tsh}\right)\widetilde{Z}\left( e^{j\omega_{k+r} \tsh}\right)\\
			-&N\left( e^{j\omega_{k+r} \tsh}\right)\widetilde{W}\left( e^{j\omega_{k+r} \tsh}\right)-L\left( e^{j\omega_{k+r} \tsh}\right)
			\Bigg\|_2^2,
		\end{aligned}
	\else
		\begin{aligned}
			\widehat{\Theta}\left(k\right) &= \arg\min_{\Theta\left(k\right)} \sum_{r=- \wsize}^{\wsize} \left\|\vphantom{\sum_{r=- \wsize}^{\wsize}}
			 D\left( e^{j\omega_{k+r} \tsh}\right)\left(\widetilde{Z}\left( e^{j\omega_{k+r} \tsh}\right)-\widehat{\widetilde{Z}}\left( e^{j\omega_{k+r} \tsh}\right)\right)\right\|_2^2 \\
			&= \arg\min_{\Theta\left(k\right)} \sum_{r=- \wsize}^{\wsize} \left\|\vphantom{\sum_{r=- \wsize}^{\wsize}}
			D\left( e^{j\omega_{k+r} \tsh}\right)\widetilde{Z}\left( e^{j\omega_{k+r} \tsh}\right)-	N\left( e^{j\omega_{k+r} \tsh}\right)\widetilde{W}\left( e^{j\omega_{k+r} \tsh}\right)-L\left( e^{j\omega_{k+r} \tsh}\right)
			\right\|_2^2,
		\end{aligned}
	\fi
\end{equation}
which has a unique closed-form solution \citep{Voorhoeve2018}.
\begin{remark}
	An unweighted version of \eqref{PFGID:eq:CostFunction} can also be minimized, either through direct non-linear optimization or by utilizing iterative reweighted methods like the Sanathanan-Koerner algorithm \citep{Sanathanan1963}. Such optimization techniques generally do not ensure convergence to a global minimizer. In addition, the weighted least-squares criterion \eqref{PFGID:eq:CostFunction} is particularly effective for practical applications \citep{Voorhoeve2018,Verbeke2020}.
\end{remark}
\begin{remark}
	\label{rem:ExcitationSignal}
	Typically, the cost function \eqref{PFGID:eq:CostFunction} has a unique closed-form solution only if the excitation signal $\widetilde{W}\left( e^{j\omega_{k+r} \tsh}\right)$ is sufficiently 'rough' within the window $r\in\mathbb{Z}_{[-\wsize,\wsize]}$ \citep{Schoukens2009}. For instance, orthogonal random-phase multisines \citep{Dobrowiecki2006} for $\widetilde{w}$ or random-phase multisines for $w_h$ meet this criterion.
\end{remark}
\ifthesismode
The frequency-lifted transfer function $\widehat{\widetilde{M}}\left( e^{j\omega_{k} \tsh}\right)$ is now identified by evaluating the unique closed-form solution \eqref{PFGID:eq:CostFunction} for all frequency bins $k\in\mathbb{Z}_{[0,N-1]}$. The developed method, where the PFG is directly evaluated through frequency-lifted time-invariant representations of multirate systems (\contributionRef{Contribution:PFGID:i}), which are identified with local modeling techniques (\contributionRef{Contribution:PFGID:ii}), is summarized in Procedure~\ref{PFGID:proc:1}.
\else
The frequency-lifted transfer function $\widehat{\widetilde{M}}\left( e^{j\omega_{k} \tsh}\right)$ is now identified by evaluating the unique closed-form solution \eqref{PFGID:eq:CostFunction} for all frequency bins $k\in\mathbb{Z}_{[0,N-1]}$. The developed method, where the PFG is directly evaluated through frequency-lifted time-invariant representations of multirate systems (C1), which are identified with local modeling techniques (C2), is summarized in Procedure~\ref{PFGID:proc:1}.
\fi
\begin{figure}[h]
	\normalsize
	\ifthesismode \else\vspace{-10pt}\fi \hrule \vspace{1.5mm}\begin{proced}[Frequency-Domain Identification of PFG through Lifting and Local Modeling] \hfill \vspace{1mm} \hrule \vspace{1mm}
		\label{PFGID:proc:1}
		\begin{enumerate}
			\item Construct excitation signal $w_h$, see \remRef{rem:ExcitationSignal}.
			\item Excite multirate system in \figRef{PFGID:fig:MRCL_GeneralizedPlant} with $w_h$ and record fast-rate performance variable $z_h$.
			\item Take DFT \eqref{PFGID:eq:DFT1} of the exogenous signal $w_h$ and performance variable $z_h$, resulting in $W_h$ and $Z_h$.
			\item Lift fast-rate signals $W_h$ and $Z_h$ into $\widetilde{W}=\mathcal{L}_f W_h$ and $\widetilde{Z}=\mathcal{L}_f Z_h$ using \eqref{PFGID:eq:fLiftedSignal}.
			\item  For frequency bins $k\in\mathbb{Z}_{[0,N-1]}$ identify the PFG $\mathcal{P}\left(e^{j\omega_k \tsh}\right)$ as follows.
			\begin{enumerate}
				\item Identify frequency-lifted closed-loop $\widehat{\widetilde{M}}\left(e^{j\omega_k\tsh}\right)$ by minimizing the local modeling cost function \eqref{PFGID:eq:CostFunction}, which has a unique global minimizer.
				\item Compute the PFG through \theoremRef{theorem:fLiftedPFG}, specifically by using \eqref{PFGID:eq:fLiftedPFGCalc}.
			\end{enumerate}
		\end{enumerate}
		\vspace{0pt} 	\hrule \ifthesismode \else\vspace{-9pt}\fi 
	\end{proced}
\end{figure}
\section{Experimental Validation}
In this section, the developed approach for direct frequency-domain identification of the intersample performance for multirate systems is experimentally validated. It is shown that the developed approach can directly identify the intersample performance in the frequency-domain through PFGs, while traditional representations or approaches cannot.
\subsection{Experimental setup}
\begin{figure}[tb]
	\centering
	\ifthesismode
		\begin{subfigure}[t]{0.45\textwidth}
		\centering
		\includegraphics{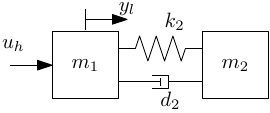}
		\caption{Sketch of experimental setup.}
	\end{subfigure}\hspace{1mm}
	\begin{subfigure}[t]{0.45\textwidth}
		\centering
		\setlength{\fboxsep}{0pt}
		\fbox{\includegraphics[width = 0.6\textwidth]{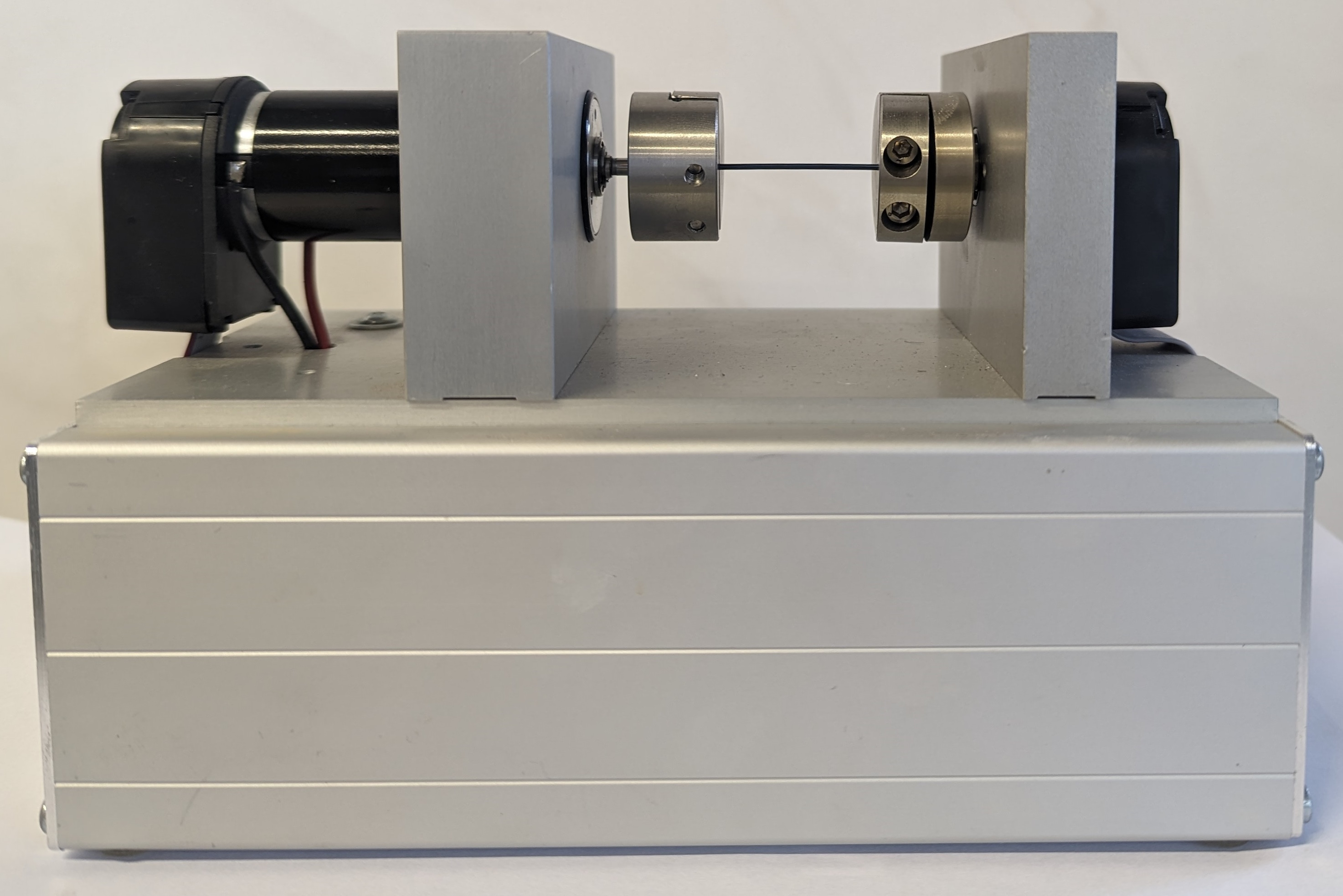}}
		\caption{Photograph of experimental setup.}
	\end{subfigure}
\else
		\begin{subfigure}[t]{0.3\textwidth}
			\centering
			\includegraphics{SchematicExperimentalSetup.pdf}
			\caption{Sketch of experimental setup.}
		\end{subfigure}\hspace{1mm}
	\begin{subfigure}[t]{0.35\textwidth}
			\centering
		\setlength{\fboxsep}{0pt}
		\fbox{\includegraphics[width = 0.6\textwidth]{PATO.jpg}}
		\caption{Photograph of experimental setup.}
	\end{subfigure}
\fi
	\caption{Experimental setup.}
	\label{PFGID:fig:ExpSetup}
\end{figure}

The developed approach is validated on the prototype motion system shown in \figRef{PFGID:fig:ExpSetup}, which consists of two rotating masses connected with a flexible shaft.
The first mass is actuated by a DC motor, and its position is measured using a incremental encoder. 

\begin{figure}[tb]
	\centering
	{\includegraphics{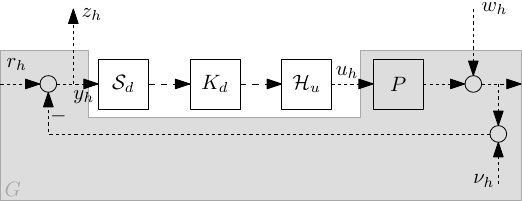}}
	\caption{Closed-loop control setup during experimental validation.}
	\label{PFGID:fig:PATOCLControlSetup}
\end{figure}

The multirate system is operating in closed-loop control as shown in \figRef{PFGID:fig:PATOCLControlSetup}, which shows the system is performing a constant velocity reference tracking task $r_h(\dt)=20\cdot 2\pi \cdot \dt$.
In addition, the output of the system $P$ is disturbed by the exogenous noise signal $w_h$, resulting in the system
\begin{equation}
	\label{PFGID:eq:ExpGeneralizedPlant}
	\begin{aligned}
		G = \left[\begin{array}{c|c}
			-1 & -P \\ \hline
			-1 & -P
		\end{array}\right]: \quad \begin{bmatrix}
		w_h \\ u_h
	\end{bmatrix} \mapsto \begin{bmatrix}
	z_h \\ y_h
\end{bmatrix}
	\end{aligned}
\end{equation}

For validation purposes, an FRF of the system is made using single-rate feedback control with a sampling rate of $\frac{\wsh}{2\pi}=\fsh=240$ Hz, where 54000 samples of the excitation signal $w_h$, input $u_h$, and output $y_h$ are used. The FRF is made in closed-loop with the indirect approach, in combination with the local rational modeling approach from \citet{McKelvey2012}, with rational degrees $R_d=R_n=R_m=3$ and window size $\wsize=150$. The FRF is seen in \figRef{PFGID:fig:PATO_FRF}. The feedback controller $K_d$ sampled at \textcolor{reviewblue}{$\frac{\wsl}{2\pi}=\fsl=80$} Hz stabilizes the downsampled system $\mathcal{S}_d P \mathcal{H}_u$ with a bandwidth of 2 Hz. Additionally, it tries to suppress any disturbance effects due to the rotational movement introduced by the reference $r_h(\dt)=20\cdot 2\pi\cdot \dt$, for example a mass imbalance. For this purpose, the loop gain is increased at 20 Hz through an inverse Notch filter. An FRF of the controller is shown in \figRef{PFGID:fig:PATOController_FRF}.
\begin{figure}[tb]
	\centering
	\ifthesismode
	\begin{subfigure}[t]{0.95\textwidth}
		\centering
		\includegraphics{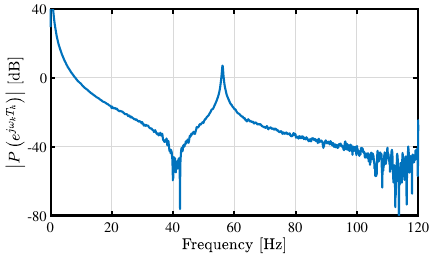}
		\caption{Validation single-rate FRF measurement of the experimental setup $P\left(e^{j\omega_k\tsh}\right)$ measured at $\fsh=240$ Hz \markerline{mblue}.}
		\label{PFGID:fig:PATO_FRF}
	\end{subfigure} \hspace{1mm}
	\begin{subfigure}[t]{0.95\textwidth}
		\centering
		\includegraphics{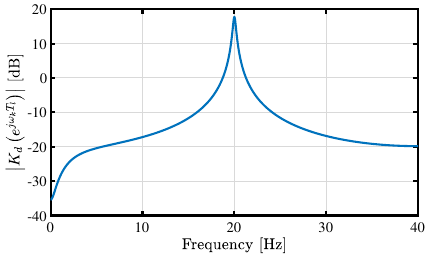}
		\caption{FRF of the feedback controller for the experimental setup $K_d\left(e^{j\omega_k\tsl}\right)$ with $\fsl=80$ Hz \markerline{mblue}.}
		\label{PFGID:fig:PATOController_FRF}
	\end{subfigure}
	\else
	\begin{subfigure}[t]{0.47\textwidth}
		\centering
		\includegraphics{PATO_FRF.pdf}
		\caption{Validation single-rate FRF measurement of the experimental setup $P\left(e^{j\omega_k\tsh}\right)$ measured at $\fsh=240$ Hz \markerline{mblue}.}
		\label{PFGID:fig:PATO_FRF}
	\end{subfigure} \hspace{1mm}
	\begin{subfigure}[t]{0.47\textwidth}
		\centering
		\includegraphics{PATOController_FRF.pdf}
		\caption{FRF of the feedback controller for the experimental setup $K_d\left(e^{j\omega_k\tsl}\right)$ with $\fsl=80$ Hz \markerline{mblue}.}
		\label{PFGID:fig:PATOController_FRF}
	\end{subfigure}
\fi
	\caption{Frequency Response Functions (FRFs) of the system $P$ and controller $K_d$.}
\end{figure}

For the experimental validation, the intersample performance of the tracking error is studied through identifying the PFG
\begin{equation}
	\label{PFGID:eq:ExpPFG}
	\begin{aligned}
			\mathcal{P}\left(e^{j \omega_k \tsh}\right)\underset{w_h \in \mathcal{W}}{=} \frac{\left\|z_h\right\|_{\mathcal{P}}}{\left\|w_h\right\|_{\mathcal{P}}} = \frac{\left\|e_h\right\|_{\mathcal{P}}}{\left\|w_h\right\|_{\mathcal{P}}}.
	\end{aligned}
\end{equation}
In addition, the PFG is compared to the slow-rate sensitivity $\mathcal{S}\left(e^{j \omega_k \tsl}\right): \: W_l\left(e^{j \omega_k \tsl}\right) \mapsto Z_l\left(e^{j \omega_k \tsl}\right)$, which is given by
\begin{equation}
	\label{PFGID:eq:ExpSlowRateSensitivity}
		\begin{aligned}
			\mathcal{S}\left(e^{j \omega_k \tsl}\right) = \left(1+K_d\left(e^{j\omega_k\tsl}\right)P_l\left(e^{j\omega_k\tsl}\right)\right)^{-1},
	\end{aligned}
\end{equation}
where $P_l$ is calculated similarly to \eqref{PFGID:eq:DownsampledFRF} using the true FRF shown in \figRef{PFGID:fig:PATO_FRF}. Further experimental settings are seen in \tabRef{PFGID:tab:expValues}.
\begin{table}[tb]
	\centering
	\caption{Experimental settings.}
	\label{PFGID:tab:expValues}
	\begin{tabular}{llll}
		\toprule
		\textbf{Variable}    & \textbf{Abbreviation} & \textbf{Value} & \textbf{Unit} \\
		\midrule
		Fast sampling frequency   & \textcolor{reviewblue}{$\fsh=\wsh/\left(2\pi\right)$}            & 240       & Hz   \\
		Slow sampling frequency & \textcolor{reviewblue}{$\fsl=\wsl/\left(2\pi\right)$}    				& 80		& Hz \\
		Downsampling factor  & \fac            & 3       & -    \\
		Number of input samples    & $N$               & 10800    & -    \\
		Number of output samples    & $M$               & 3600    & -    \\
		\bottomrule
	\end{tabular}
\end{table}
\ifthesismode
\FloatBarrier
\fi
\subsection{Intersample Analysis of the Experimental Setup}
\label{sec:ExpIntersampleAnalysis}
In this section, it is shown that traditional slow-rate FRFs or the on-sample behavior cannot accurately represent the intersample behavior of an experimental multirate system, in contrast to the PFG \eqref{PFGID:eq:ExpPFG}.

\begin{figure}[tb]
	\centering
	\ifthesismode
	\includegraphics{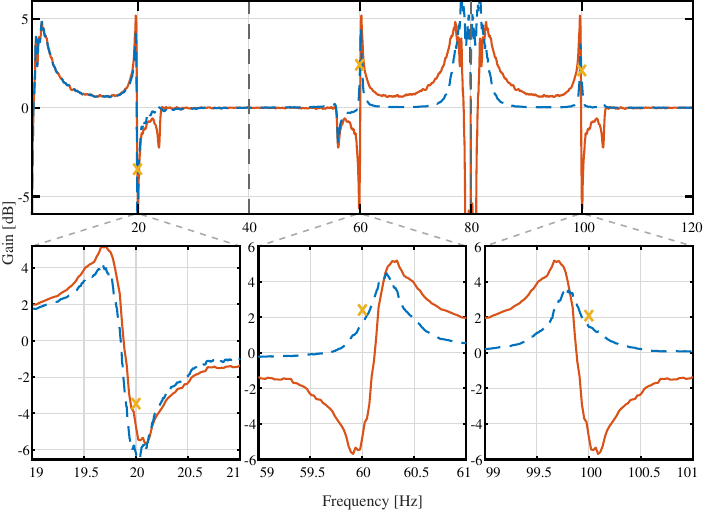}
	\else
	\includegraphics{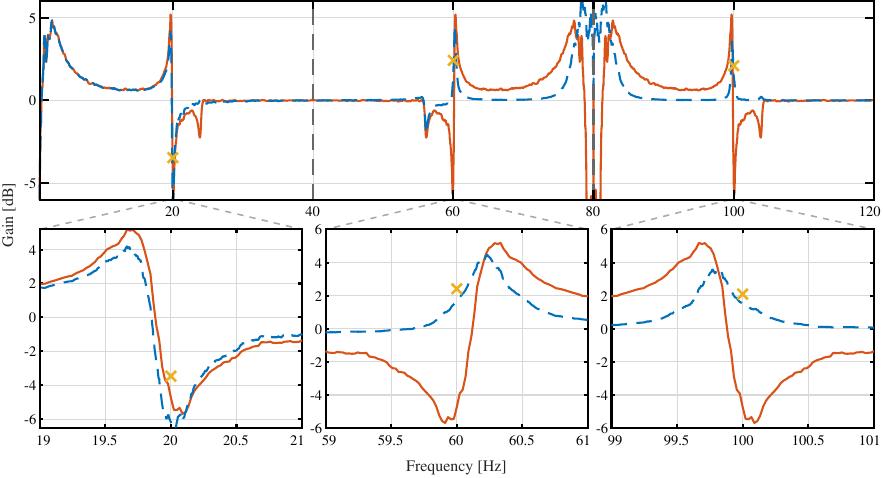}
	\fi
	\caption{The slow-rate sensitivity $\mathcal{S}\left(e^{j \omega_k \tsl}\right)$ \eqref{PFGID:eq:ExpSlowRateSensitivity} \markerline{mred} cannot asses intersample performance, unlike the PFG $\mathcal{P}\left(e^{j \omega_k \tsh}\right)$ derived with the validation FRF and \eqref{PFGID:eq:AnalyticalPFG} \markerline{mblue}[densely dashed]. For instance, it deviates over a factor 2 (6 dB) at 60 and 100 Hz from the PFG. The PFG is validated for several single-sinusoidal measurements \markerline{myel}[solid][x][3][1][0].} 
	\label{PFGID:fig:ExpPFGVsSens}
\end{figure}

First, it is shown that the slow-rate sensitivity does not represent the frequency-domain intersample performance by comparing it to the PFG. The PFG is computed analytically using \citet[Lemma~4]{Oomen2007} and a fast-rate validation FRF $P\left(e^{j\omega_k\tsh}\right)$ measured at $\fsh=240$ Hz. Specifically, the PFG is computed as
\begin{equation}
	\label{PFGID:eq:AnalyticalPFG}
	\begin{aligned}
		\mathcal{P}\left(e^{j\omega_k\tsh}\right) = \sqrt{\sum_{f=0}^{\fac-1}\Big|c_f\left(e^{j\omega_k\tsh}\right)\Big|^2},
	\end{aligned}
\end{equation}
where $c_f\left(e^{j\omega_k\tsh}\right) \in\mathbb{C}$ is given by
\begin{equation}
	\ifthesismode
	\resizebox{\linewidth}{!}{$\displaystyle
	\else
	\fi
	\begin{aligned}
	c_f\left(e^{j\omega_k\tsh}\right) =	 \begin{cases}
			G_{11}\!\left(e^{j\omega_k\tsh}\right)\!+\!\frac{1}{\fac}G_{12}\!\!\left(e^{j\omega_k\tsh}\right)\mathcal{I}_{\scriptscriptstyle ZOH}\!\left(e^{j\omega_k\tsh}\right)
			\cdot Q_d\left(e^{j\omega_k\tsl}\right)G_{21}\left(e^{j\omega_k\tsh}\right),		& f=0, \\[10pt]
			\frac{1}{\fac}G_{12}\!\left(e^{j\omega_k\tsh}\phi^f\right)\mathcal{I}_{\scriptscriptstyle ZOH}\!\left(e^{j\omega_k\tsh}\phi^f\right) Q_d\left(e^{j\omega_k\tsl}\right)G_{21}\left(e^{j\omega_k\tsh}\right),		& f\neq 0,
	\end{cases}
	\end{aligned} \ifthesismode
$}
\else
\fi
\end{equation}
with frequency shift $\phi=e^{j2\pi/\fac}=e^{j\wsh\tsh/\fac}$, and $G_{ij}$ is defined in \eqref{PFGID:eq:ExpGeneralizedPlant} for the experimental setup. Both the slow-rate sensitivity \eqref{PFGID:eq:ExpSlowRateSensitivity} and the PFG, which is calculated using \eqref{PFGID:eq:AnalyticalPFG} and the validation FRF in \figRef{PFGID:fig:PATO_FRF}, are shown in \figRef{PFGID:fig:ExpPFGVsSens}.

\begin{figure}[tb]
	\centering
	\includegraphics{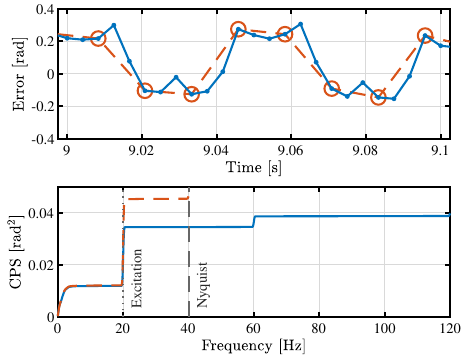}
	\caption{For excitation of 20 Hz, the on-sample performance $z_l$ \markerline{mred}[densely dashed][o][2][0.85] is relatively similar to the intersample performance $z_h$ \markerline{mblue}[solid][*][1] (top). The on-sample CPS$_l$ \eqref{PFGID:eq:PFG_CPS} \markerline{mred}[densely dashed] and intersample CPS$_h$ \eqref{PFGID:eq:PFG_CPS} \markerline{mblue} show that they both have a similar dominant component at 20 Hz (bottom).}
	\label{PFGID:fig:ExpSingleSinusoid20}
\end{figure}

\begin{figure}[tb]
	\centering
	\ifthesismode
	\begin{subfigure}[t]{0.95\textwidth}
		\centering
		\includegraphics{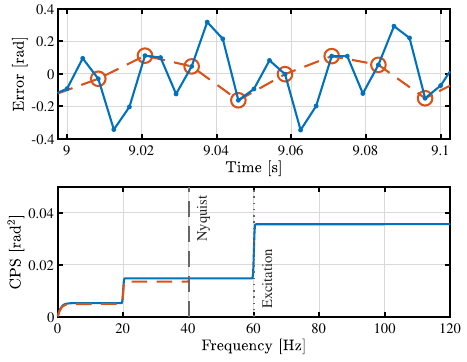}
		\caption{Excitation of 60 Hz.}
	\end{subfigure}
	\hspace{1mm}
	\begin{subfigure}[t]{0.95\textwidth}
		\centering
		\includegraphics{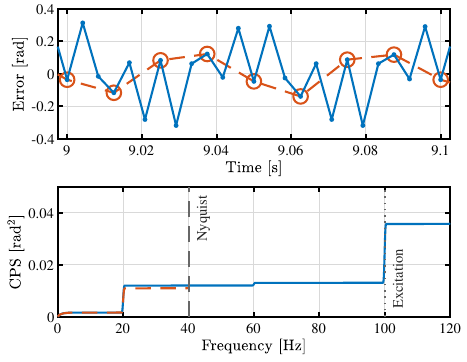}
		\caption{Excitation of 100 Hz.}
	\end{subfigure}
	\else
	\begin{subfigure}[t]{0.48\textwidth}
		\centering
		\includegraphics{ExperimentalPerformance60}
		\caption{Excitation of 60 Hz.}
	\end{subfigure}
	\hspace{1mm}
	\begin{subfigure}[t]{0.48\textwidth}
		\centering
		\includegraphics{ExperimentalPerformance100}
		\caption{Excitation of 100 Hz.}
	\end{subfigure}
\fi
	\caption{When exciting the system beyond the Nyquist frequency, the intersample performance $z_h$ \markerline{mblue}[solid][*][1] is significantly worse than the on-sample performance $z_l$ \markerline{mred}[densely dashed][o][2][0.85] (top). The intersample CPS$_h$ \eqref{PFGID:eq:PFG_CPS} \markerline{mblue} is significantly higher than the on-sample CPS$_l$ \eqref{PFGID:eq:PFG_CPS} \markerline{mred}[densely dashed], since the dominant components beyond the Nyquist frequency are not observed by the on-sample CPS$_l$ (bottom).}
	\label{PFGID:fig:ExpSingleSinusoidBeyond}
\end{figure}

Second, the difference in the slow-rate sensitivity and the PFG is illustrated through performing three time-domain validation experiments by applying single-sinusoidal disturbances having frequencies 20, 60, and 100 Hz to $w_h$, which show a different between on-sample and intersample performance. Note that for slow-rate sampling $w_l=\mathcal{S}_dw_h$ these signals have the same frequency of 20 Hz.
The Cumulative Power Spectrum (CPS) of the fast-rate and slow-rate performance variables are respectively defined as
\begin{equation}
	\label{PFGID:eq:PFG_CPS}
	\begin{aligned}
		\text{CPS}_h\left(e^{j\omega_k\tsh}\right) = \sum_{i=0}^{k} Z_h^\prime\left(e^{j\omega_i\tsh}\right)f_{r,h}, && 
		\text{CPS}_l\left(e^{j\omega_k\tsl}\right) = \sum_{i=0}^{k} Z_l^\prime\left(e^{j\omega_i\tsl}\right)f_{r,l},
	\end{aligned}
\end{equation}
where the power spectral densities $Z_h^\prime\left(e^{j\omega_i\tsh}\right) $ and $Z_l^\prime\left(e^{j\omega_i\tsl}\right)$ can for example be determined using Welch's method \citep{Welch1967} with frequency resolutions $f_{r,h}$ and $f_{r,l}$. The performance variables and their CPS determined with Welch's method for applying the three single-sinusoidal disturbances are shown in \figRef{PFGID:fig:ExpSingleSinusoid20} and \figRef{PFGID:fig:ExpSingleSinusoidBeyond}.

From the slow-rate sensitivity and the PFG in \figRef{PFGID:fig:ExpPFGVsSens}, in addition to the performance variables for single-sinusoidal disturbances in \figRef{PFGID:fig:ExpSingleSinusoid20}, and \figRef{PFGID:fig:ExpSingleSinusoidBeyond}, it becomes clear that the slow-rate sensitivity and the on-sample behavior cannot accurately represent the intersample behavior of the multirate system, which is concluded from the following observations.
\begin{itemize}
	\item While the PFG and the slow-rate sensitivity in \figRef{PFGID:fig:ExpPFGVsSens} are relatively similar below 20 Hz, for the intersample performance beyond 20 Hz they are significantly different.
	\item While the on-sample performance in \figRef{PFGID:fig:ExpSingleSinusoid20} and \figRef{PFGID:fig:ExpSingleSinusoidBeyond} is good, supported by the suppression of -3.17 dB of the slow-rate sensitivity in \figRef{PFGID:fig:ExpPFGVsSens}, the intersample performance in \figRef{PFGID:fig:ExpSingleSinusoidBeyond} deteriorates significantly for 60 and 100 Hz. Specifically, the root mean squared intersample performance deteriorates respectively by factors 1.63 and 1.79 compared to the on-sample performance.
\end{itemize}

\ifthesismode
\FloatBarrier
\fi
\subsection{Experimental Identification of PFG}
\begin{figure}[tb]
	\centering
	\ifthesismode
	\includegraphics{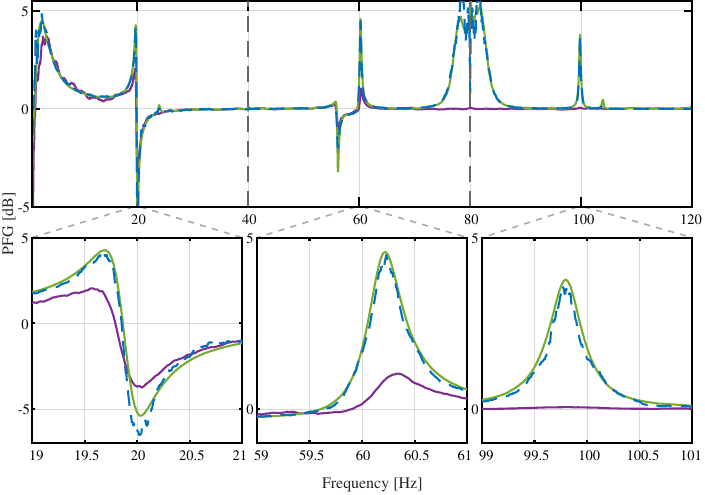}
	\else
	\includegraphics{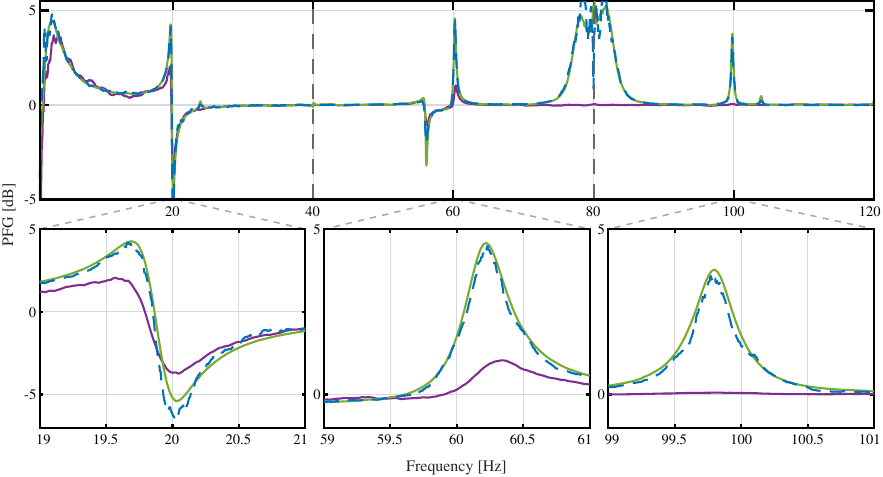}
	\fi
	\caption{The true PFG determined with the validation FRF and \eqref{PFGID:eq:AnalyticalPFG} \markerline{mblue}[densely dashed] is accurately identified by the developed approach \markerline{mgreen} using \theoremRef{theorem:fLiftedPFG}, in contrast to a direct approach that neglects the multirate behavior \markerline{mpurple}.}
	\label{PFGID:fig:ExpEstPFG}
\end{figure}

In this section, the PFG is directly and accurately identified in a single identification experiment for the experimental setup using the developed method. In addition, the developed method is compared to an approach that neglects the multirate behavior by assuming single-rate sampling. Therefore, the compared methods are the following.
\begin{itemize}
	\item An approach that neglects the multirate behavior by identifying the closed-loop $w_h\mapsto z_h$ directly through local rational modeling.
	\item The developed approach that identifies the frequency-lifted system $\widetilde{M}$ and calculates the PFG using \eqref{PFGID:eq:fLiftedPFGCalc}.
\end{itemize}
Both methods utilize rational degrees $R_d=R_n=R_m=3$ and window size $\wsize=60$. Additional settings during experimentation are shown in \tabRef{PFGID:tab:expValues}. The methods are compared to the validation PFG computed in \secRef{sec:ExpIntersampleAnalysis}. The identified PFG by both methods is shown in \figRef{PFGID:fig:ExpEstPFG}.
The PFG is accurately identified in a single identification experiment using the developed approach, while the approach that neglects the multirate behavior is not able to accurately identify the PFG as shown in \figRef{PFGID:fig:ExpEstPFG}. 

In conclusion, the observations show that the PFG is essential for quantifying the intersample performance of a multirate or sampled-data system, which is accurately and directly identified by the developed approach. \textcolor{reviewblue}{The} approach that neglects the multirate behavior does not represent the full intersample behavior \textcolor{reviewblue}{of the experimental system, because it does not accurately identify its PFG}.
\ifthesismode
\FloatBarrier
\fi

\section{Conclusions}
The results in this \manuscript enable direct single-experiment identification of frequency-domain intersample performance in closed-loop multirate systems. The PFG, which is a frequency-domain representation for sampled-data or multirate systems, is directly evaluated using the frequency-lifted system, which is a time-invariant representation of the multirate system. The multiple aliased frequencies in the multivariable time-invariant representation are effectively disentangled through multivariable local modeling techniques. The time-invariant representation is directly identified in a single identification experiment. For an experimental prototype motion system it is shown that slow-rate or on-sample FRFs cannot be used to analyze the intersample performance of a multirate system. In sharp contrast, the PFG represents the intersample performance of the experimental system in the frequency domain, which is accurately and directly identified in a single identification experiment by the developed approach. Therefore, the developed method is a key enabler for sampled-data and multirate control by directly identifying the performance of digital control systems in the frequency-domain, including the intersample performance.

\bibliographystyle{cas-model2-names}
\bibliography{../../library,../../libraryMaths}

\end{document}